\documentstyle[aps,prl,epsfig,multicol]{revtex}
\begin{document}

\newsavebox{\GSIM}
\sbox{\GSIM}{\raisebox{-1ex}{$\ \stackrel{\textstyle>}{\sim}\ $}}
\newcommand{\gsim}{\usebox{\GSIM}}

\title{Exciting Relative Number Squeezed Particles from Condensates Using Stimulated Light Scattering}
\author{D.C.~Roberts, T.~Gasenzer and K.~Burnett}
\address{Clarendon Laboratory, Department of Physics, University of Oxford,
Oxford OX1~3PU, United Kingdom}
\date{\today}
\maketitle

\begin{abstract}
We investigate the relative particle number squeezing produced in the excited states of a weakly interacting condensate at zero temperature by stimulated light scattering using a pair of lasers.  We shall show that a modest number of relative number squeezed particles can be achieved when atoms with momentum $k$, produced in pairs through collisions in the condensate, are scattered out by their interaction with the lasers. This squeezing is optimal when the momentum $k$ is larger than the inverse healing length, $k>k_0$.  This modest number of relative number squeezed particles has the potential to be amplified in four-wave-mixing experiments.   
\end{abstract}

\begin{multicols}{2}

A macroscopic source of strongly entangled atoms would be immensely valuable for such applications as Heisenberg limited clocks \cite{clock}, quantum information processing (for reviews see \cite{qc}), and tests of quantum state reduction \cite{penrose}. 
Various methods of producing macroscopic entangled atoms using Bose-Einstein condensates (BECs) have been proposed, such as four-wave-mixing \cite{4wave} and utilizing spin-exchanging collisions in a trap\cite{spinexchange1,spinexchange2}. The goal of this letter is to investigate the relative number squeezing produced in excited states of a condensate using stimulated light scattering \cite{gasenzerref}.  We shall briefly discuss how this relative number squeezing can be converted into entanglement.   The main result of this letter is that if atoms, produced in pairs through collisions in the condensate, are scattered out by their interaction with the lasers, then a modest number of almost optimally relative number squeezed particles can be produced (as shown in Figure 1).  

We consider a homogeneous condensate of atoms whose collisional interaction is described by a contact interparticle pseudopotential $\Phi (r)=U_0 \delta^{(3)}(r)$.
Relative number squeezing in the situation considered in this paper originates from pair producing collisions in a dilute BEC.  To understand the nature of these pair-producing collisions present in a dilute uniform BEC, we first consider how weak repulsive interactions affect the ground state. Correlations between free particle states with momenta $k$ and $-k$ (all momenta are understood to be vectors) are present in the condensate ground state, $|\phi_0 \rangle$. These arise from the momentum conserving interaction between two condensate atoms colliding that admix momentum states $k$ and $-k$ into the ground state.  Observation of the suppression of the static structure factor at low $k$ \cite{ketterle} is evidence for the existence of these correlated pair excitations. In general this homogeneous model should be a good approximation for the large condensates as produced in recent experiments.

In this paper, we investigate the possibility of making these correlations more accessible by transferring one particle of such a pair into a state of higher momentum, e.g.~$k$ into $k+\Delta k$, using stimulated light scattering. We find that squeezing of the population difference between the modes $-k$ and $k+\Delta k$ in the final state may be achieved in principle only for momenta $k$ on the order of or larger than the inverse healing length \cite{heal}. 

In the homogeneous model condensate the eigenmodes of the weakly interacting Hamiltonian, which we define below, are travelling plane waves. Furthermore, we impose periodic boundary conditions in a finite volume $V$, which  makes the momentum spectrum discrete. This should be a reasonable approximation since one can fix the photon momenta by tuning the two lasers and the atomic momenta by assuming a measurement or an extraction mechanism to be concentrated on the relevant momenta.  

We can describe the condensate with a second quantized Hamiltonian of the form: 
\begin{equation}
\hat H = \sum_{k} \frac{ \hbar^2 k^2}{2 m} \hat a_k^\dag \hat a_k+\frac{U_0}{2 V} \sum_{k,k_1,k_2} \hat a_k^\dag \hat a_{k_1}^\dag \hat a_{k_2} \hat a_{k+k_1-k_2},
\end{equation}
where $U_0=4 \pi \hbar^2 a/m$, $a$ is the 2-body scattering length that characterizes the collisions between the atoms, and $m$ is the atomic mass.  

 By assuming that there is a macroscopic  population in the $k=0$ momentum state and neglecting terms of order $N^{1/2}$,  where $N$ is the total number of bosons, we can approximate this Hamiltonian to describe a dilute condensate at low temperature \cite{fetter} obtaining
\begin{eqnarray}
\hat H_{BEC}  &\approx&  \sum_{k \ne 0} \left( \frac{ \hbar^2 k^2}{4 m} + \frac{N U_0}{V} \right) (\hat a_k^\dag \hat a_k+\hat a_{-k}^\dag \hat a_{-k)}  \nonumber \\  
 && +\ \frac{N U_0}{V}(\hat a_k \hat a_{-k}+\hat a_k^\dag \hat a_{-k}^\dag).
\end{eqnarray}
 Note that we have redefined the ground state energy to be zero.  We then diagonalize $\hat H_{BEC}$  by introducing a linear transformation, $\hat a_k=u_k \hat \alpha_k-v_k \hat \alpha_{-k}^{\dag}$, from particle operators, $\hat a_k$, to quasiparticle operators, $\hat \alpha_k$, where   $u_k=(1-\beta_k^2)^{-1/2}$, $v_k=\beta_k(1-\beta_k^2)^{-1/2}$, $\beta_k=1+y_k^2-y_k(2+y_k^2)^{1/2}$, $y_k=|k|/|k_0|$, $k_0^2=8 \pi a N_0/V$, and $N_0$ is the number of particles in the  $k=0$ momentum state. We can write
\begin{equation}
\label{diag}
\hat H_{BEC} =    \sum_{k \ne 0} \hbar \omega_k \hat \alpha_k^\dag \hat \alpha_k,
\end{equation}
where $\omega_k=(\hbar k_0^2/2m)y_k\sqrt{2+y_k^2}$ \cite{heal}.
  Throughout this paper we are assuming that our system is at, or modestly perturbed from, a quasiparticle vacuum or $T=0$.  The ground state $|\phi_0 \rangle$, for the uniform weakly interacting BEC, is then defined by $\hat \alpha_k | \phi_0 \rangle=0$ for $k \ne 0$, implying
\begin{equation}
\label{gs}
|\phi_0 \rangle = \sum_{n=0}^\infty \sum_{k} \left( -\frac{v_k}{u_k} \right)^n|n_{-k},n_{k} \rangle.
\end{equation}
This state has the form of a set of two-mode squeezed states \cite{Barnett} and consequently has the property of being perfectly  relative number squeezed, or having zero variance in the relative number of particles in $k$ and $-k$, i.e. $[\Delta (n_k-n_{-k})]^2=\langle (\hat n_k-\hat n_{-k})^2 \rangle-\langle \hat n_k-\hat n_{-k}\rangle^2=0$. 

Squeezing in the relative number of particles between states is not equivalent to entanglement, but these squeezed states, if not already entangled, can be converted into entangled states.  For example, a twin Fock state, which is an optimally relative particle number squeezed system, can be converted into an entangled system via a 50/50 beam-splitter \cite{Barnett2,Braunstein,Dunningham01}.   The ground state of a dilute uniform BEC (eq.~(\ref{gs})) is optimally relative number squeezed.  However, this is presumably of practical use only if the populations of these squeezed states can be separated from the condensate ground state.   

We shall quantify the degree of relative number squeezing by defining the squeezing parameter, $\xi_{k,k'}$, to be
\begin{equation}
\xi_{k,k'} \equiv \frac{[\Delta (n_k-n_{k'})]^2}{ n_k+n_{k'} } 
\end{equation}
where $  n_k= \langle \hat n_k \rangle=\langle \hat a^\dag_k \hat a_k \rangle $ is the number of particles in a momentum state $k$. This squeezing parameter is equivalent to the one introduced by \cite{wineland} in the spin-squeezing formalism to take into account the correlations of particles discussed by \cite{kitagawa}.  If the two relevant states are independent and coherent then $\xi_{k,k'}=1$.  On the other hand, for optimally relative number squeezed populations of the momentum states $k$ and $k'$, $[\Delta (n_k-n_{k'})]^2=0$ implying $\xi_{k,k'}=0$.  So as $\xi_{k,k'}$ decreases, the relative number squeezing between the two states improves.

Relative particle number squeezed systems can be a valuable source for entanglement if the particles in the squeezed states can be extracted from the condensate.   In this letter, we investigate how to accomplish this extraction using stimulated light scattering.  In this method a condensate is illuminated using two laser beams with a wave vector difference of $\Delta k$ and a frequency difference of $\delta$ \cite{bragg}.  Since we are approximating quasiparticles by travelling plane waves, the outcome of a location measurement of particles is completely undetermined, and the scattering strictly speaking does not lead to an extraction, where an atom moves out along a defined trajectory.  However, since we are interested only in populations, we assume that particles with a momentum large enough to leave a trapped condensate may be described approximately by plane waves. In a more realistic formulation the particles would be described by wave packets whose momentum distribution is smeared out around the momentum of this plane wave.  

In the remaining part of this letter, we study ways to obtain the best possible amount of squeezing between atomic states of different momenta.
We calculate the time evolution of the system and the relevant populations using perturbation theory. We treat the effective coupling between the laser and the condensate, $\widetilde \Omega$, as a perturbation to the Hamiltonian (\ref{diag}) describing a weakly interacting condensate.  The results presented below have been calculated to the order $\widetilde \Omega^2$. Hence, we are implicitly assuming that the lasers do not drive the system far out of equilibrium, so that the Bogo\-liu\-bov approximation still holds.

The effective Hamiltonian can be written as  $\hat H = \hat H_{BEC} + \hat H_{L}$, where
\begin{equation}
\hat H_{L}=\mbox{$\frac{1}{2}$}\hbar \widetilde \Omega \sum_k(\hat a^\dag_k \hat a_{k-}e^{-i \delta t}+\mbox{h.c.})
\end{equation}
 (subscripts $k_\pm$ are shorthand for $k \pm \Delta k$).  Using the quasi-particle transformation already introduced, the Bogliubov approximation implying $\hat a_0,\hat a^\dag_0 \rightarrow \sqrt{N_0}$, and noting that $(\hat a^\dag_{\Delta k}+ \hat a_{-\Delta k})=(u_{\Delta k}-v_{\Delta k})(\hat \alpha^\dag_{\Delta k}+ \hat \alpha_{-\Delta k})$ we obtain
\begin{eqnarray}
\hat H_{L}&=&\mbox{$\frac{1}{2}$}\hbar \widetilde \Omega e^{-i \delta t}\Big[\sqrt{N_0}(u_{\Delta k}-v_{\Delta k})(\hat \alpha_{\Delta k}^\dag+\hat \alpha_{-\Delta k})
\nonumber\\
&& +\ \sum_{k \ne 0,\Delta k} \Big(u_{k k_-} \hat \alpha_k^\dag \hat \alpha_{k_-}-u_{k_-}v_k\hat \alpha_{-k} \hat \alpha_{k_-}
\nonumber\\
&&\qquad\qquad -\ u_k v_{k_-}\hat \alpha_k^\dag \hat \alpha^\dag_{-k_-}\Big)\Big] +\mbox{h.c.}
\end{eqnarray}
 where $u_{k k'}=u_k u_{k'}+v_k v_{k'}$, $v_{k k'}=u_k v_{k'}+v_k u_{k'}$. Note the factor of $(u_{\Delta k}-v_{\Delta k})$ originates because of the destructive interference between two resonant paths connecting the $k=0$ momentum state to the momentum states $\pm \Delta k$ \cite{ketterle}. We have approximated the internal structure of the atom to be represented by a two-level system.  The coupling is given by $\widetilde \Omega =|\Omega|^2/2\Delta $, where $\Omega$ is Rabi frequency of the one-photon transition between the internal atomic states. The detuning, $\Delta$, of the laser from the internal atomic transition frequency is assumed to be sufficiently large to adiabatically eliminate the intermediate state. We have neglected the AC-Stark shifts.

\begin{figure}[ht]
\begin{center}
\epsfig{file={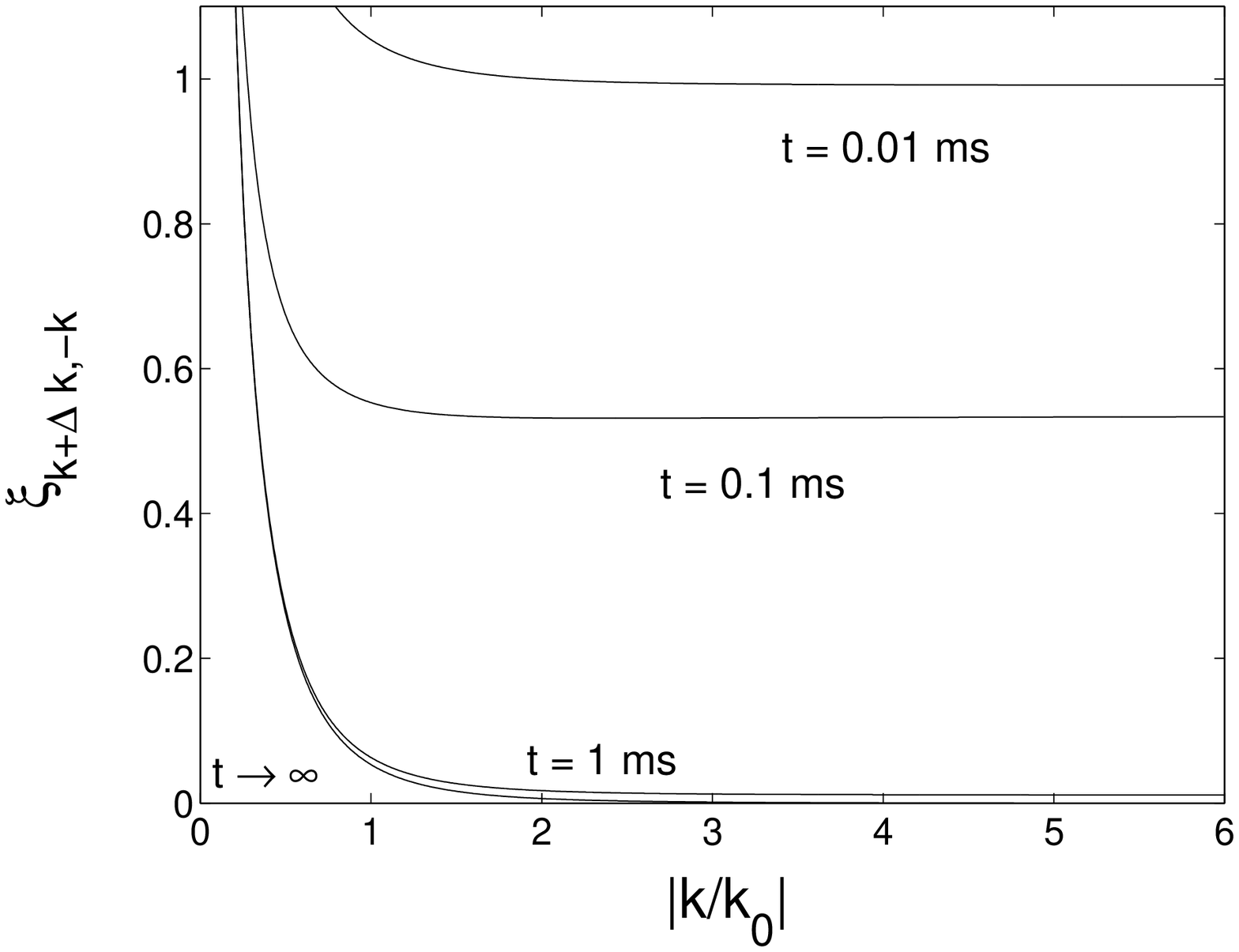},height=5cm,width=7cm,angle=0}\\[0.3cm]
\caption{\label{xisimp} The squeezing parameter $\xi_{k+\Delta k,- k}$, for $\Delta k=k/2$, as a function of $|k|$ for successive times for typical experimental parameters: $N_0 =10^7$ $^{23}$Na atoms, Volume $V=10^{-7}\,$cm$^3$, $a=2.8\,$nm, Rabi frequency $\Omega= (2\pi)\,1.8\,$MHz, and detuning $\Delta= (2\pi)\,1\,$GHz implying $n_0=10^{14}\,$cm$^{-3}$, $\hbar k_0^2/2m = (2\pi)\,1.5\,$kHz and  $ \widetilde \Omega/(\hbar k_0^2/2m) = 1.0$.  Note that the squeezing improves with increasing time and increasing momentum.}
\end{center}
\end{figure}

We investigate the squeezing behavior of particle populations of modes $k \ne \pm \Delta k$ that are not directly coupled to the condensate mode ($k=0$) by their interaction with the lasers.  We find that the best squeezing occurs between populations of momentum states $k+\Delta k$ and $-k$ as the two laser beams with their frequency difference tuned to $\delta=  \omega_k +\omega_{k+\Delta k}$ excite a particle from a correlated pair in the ground state \cite{similar}. In this case the interaction Hamiltoninan is $\hat H_{L} \approx -\frac{1}{2} \hbar \widetilde \Omega\, v_{kk_+} \hat \alpha^\dag_{k+\Delta k} \hat \alpha_{-k}^\dag+\,$h.c.,  which has the form of a two-mode squeezing operator.  Therefore, the {\it quasi}particle populations at momenta $k+\Delta k$ and $-k$ are optimally relative number squeezed to the order of the calculation.  Since this simplified Hamiltonian is proportional to $v_{kk_+}$, it is clear that the collisions during the stimulated photon scattering are the essential process for the production of squeezing.  
We project these optimally squeezed quasiparticle states onto a particle basis to obtain the particle squeezing parameter $\xi_{k+\Delta k,-k}$ as a function of $|k|$ (assuming $\Delta k=k/2$) for successive times as seen in Fig.~\ref{xisimp}.  
As time increases the squeezing improves and the squeezing parameter asymptotically approaches a minimum value   
\begin{eqnarray}
\xi_{k+\Delta k,-k} &\approx& [(u_{k_+}^2+v_{k_+}^2)u_{k_+}^2+(u_{k}^2+v_{k}^2)u_{k}^2 
\nonumber\\
&&\ -\ 2 u_k^2 u_{k_+}^2]/[u_{k_+}^2+u_{k}^2].
\end{eqnarray}  
Although perturbation breaks down at large times, the limit of the validity of our perturbative calculation is very close to the asymptotic limit.  On resonance, the particle populations of atoms with momenta $k+\Delta k$ and $-k$ are approximately given by
\begin{eqnarray}
n_{k+\Delta k} &\approx& v_{k_+}^2+ \mbox{$\frac{1}{4}$}(\widetilde \Omega t)^2 u_{k_+}^2 v_{kk_+}^2,\\
n_{-k} &\approx& v_{k}^2+\mbox{$\frac{1}{4}$}(\widetilde \Omega t)^2 u_{k}^2 v_{kk_+}^2,
\end{eqnarray}
respectively.
 We can see from Figure \ref{xisimp} that the relevant states are approximately relative number squeezed ($\xi_{k+\Delta k,-k}$ close to zero) in the particle regime $|k|>k_0$.  For example, if we assume the same experimental parameters as used in Figure \ref{xisimp}, the two lasers' combined pulse with a duration of  $10\,$ms should produce roughly $60$ particles in both momentum states  $3 k_0$ and $-2 k_0$.  These momentum states should be almost optimally squeezed with $\xi_{k+\Delta k,-\Delta k} \approx 0$. 
\begin{figure}[ht]
\begin{center}
\epsfig{file={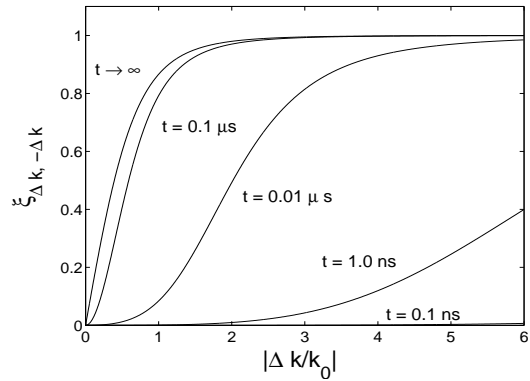},height=5cm,width=7cm,angle=0}\\[0.2cm]
\caption{\label{squeezemac} The squeezing parameter $\xi_{\Delta k,-\Delta k}$ as a function of $|\Delta k|$ for successive times for the same experimental parameters as used in Figure 1.  Note that the squeezing deteriorates with increasing time and the squeezing is concentrated at the low momentum ($|\Delta k|<k_0$) states. }
\end{center}
\end{figure}
It is also instructive to investigate the leading order process in which particles in the condensate mode $k=0$ are directly scattered into modes $\pm\Delta k$.  We consider the relative particle number squeezing between the momentum states $k=\Delta k$ and $k=-\Delta k$.  Assuming the laser beams' frequency difference is tuned to the resonance $\delta = \omega_{\Delta k}$, the interaction Hamiltonian is $\hat H_{L} \approx \frac{1}{2}\hbar \widetilde \Omega\, \sqrt{N_0} (u_{\Delta k}-v_{\Delta k})\,\hat \alpha_{\Delta k}^\dag +\,$h.c., so there are no correlations between quasiparticles to the order of our calculation and therefore no relative number squeezing between quasiparticle populations.  However, when we project the excited quasiparticle states onto a particle basis there is significant relative number squeezing in the phonon (or quasiparticle) regime, $|k|<k_0$, \cite{kmin} (as shown in Figure \ref{squeezemac}).
This squeezing originates from the initial state (eq. (\ref{gs})) and it is clear that the squeezing deteriorates when additional atoms are scattered from the condensate into the modes $\pm\Delta k$. This is contrary to the previous situation where relative particle number squeezing originates from the collisions during the scattering process and improves with time.

In summary, we have considered a situation where one can excite relative number squeezed particles using stimulated light scattering.  We have shown that by considering states that are not directly coupled to the macroscopically populated $k=0$ state, one can in principle excite and extract populations of strongly relative number squeezed particles.  We expect that the dominant process to limit this relative number squeezing will be the rescattering of atoms in collisions before they leave the condensate.  Assuming a cubic volume $V$, we estimate the percentage of rescattered particles to be smaller than  $r=\sigma n_0 V^{1/3}=42\%$. We expect that the inclusion of such processes into our calculation would yield an upper bound for the achievable squeezing and an approximate time when this optimum is reached. Another measure of the loss is the mean decay time due to quasiparticle collisions. At zero temperature the main loss mechanism is given by Beliaev damping processes, in which either a quasiparticle collides with a condensate atom to produce a pair of quasiparticles or two quasiparticles collide and thereby produce an atom in the condensate and another quasiparticle. For $|k|\gg k_0$, the inverse Beliaev decay width $\tau_{k}=\gamma_{k}^{-1}=m/(8\pi a^2n_0\hbar k)$ \cite{Beliaev58} which may also be obtained from kinetic theory, is for our above example $\tau_{2k_0}\simeq3.5\,$ms, giving an estimate for the time when deteriorating processes become significant.

The particles produced in the relative number squeezed modes might be used as seeds for four-wave-mixing experiments. Since the introduction of uncorrelated particles can be minimized, the squeezing in the output of such experiments can be greatly improved. Note that a similar amplification method has recently been demonstrated \cite{LamasLinares01} for the production polarization-entangled photons, using stimulated emission inside a nonlinear medium. 

This work was financially supported by the Marshall Trust (D.C.R.), the A.v.~Humboldt-Foundation and a Marie Curie Fellowship of the European Community, contract no.~HPMF-CT-1999-0023 (T.G.), and the United Kingdom Engineering and Physical Sciences Research Council.

\end{multicols}

\begin{thebibliography}{9}
\bibitem{gasenzerref}  
In \cite{gasenzer},  we have investigated the squeezing behavior between photons and atoms coupled out from a BEC.
\bibitem{heal} 
$1/k_0$ is generally referred to as the healing length. $k_0$ separates the phonon-type- from the free-particle-type energy spectrum. In the homogeneous case $k_0^2=8\pi an_0$, where $a$ is the scattering length and $n_0$ the condensate density.
\bibitem{bragg} 
Note that this stimulated light scattering is called Bragg diffraction for times long enough that there is a substantial population in the diffracted mode, $t>m/(\hbar|\Delta k|^2)$.
\end{thebibliography}
\end{document}